\documentclass[aps,prc,preprint,superscriptaddress,showpacs,showkeys]{revtex4-1}
\usepackage{graphicx}
\begin{document}
\title{Giant Dipole Resonance studies in Ba isotopes at $\textbf{ {E/A}}\sim$ 5 MeV}
\author{C.~Ghosh}
\affiliation{Department of Nuclear and Atomic Physics, Tata Institute of Fundamental Research, Mumbai  400005, India.}
\author{A.K.~Rhine Kumar}
\affiliation{Department of Nuclear and Atomic Physics, Tata Institute of Fundamental Research, Mumbai  400005, India.}
\author{Balaram~Dey}
\affiliation{Department of Nuclear and Atomic Physics, Tata Institute of Fundamental Research, Mumbai  400005, India.}
\author{V.~Nanal}
\email[]{nanal@tifr.res.in}
\affiliation{Department of Nuclear and Atomic Physics, Tata Institute of Fundamental Research, Mumbai 400005, India.}
\author{R.G.~Pillay}
\affiliation{Department of Nuclear and Atomic Physics, Tata Institute of Fundamental Research, Mumbai 400005, India.}
\author{P.~Arumugam}
\affiliation{Department of Physics, Indian Institute of Technology, Roorkee 247667, India.}
\author{K.V.~Anoop}
\affiliation{Pelletron Linac Facility, Tata Institute of Fundamental Research, Mumbai  400005, India.}
\author{N.~Dokania}
\affiliation{Department of Nuclear and Atomic Physics, Tata Institute of Fundamental Research, Mumbai 400005, India.}
\author{Abhijit~Garai}
\affiliation{India-based Neutrino Observatory, Tata Institute of Fundamental Research, Mumbai  400005, India.}
\author{Ghnashyam Gupta}
\affiliation{Department of Nuclear and Atomic Physics, Tata Institute of Fundamental Research, Mumbai 400005, India.}
\author{E.T.~Mirgule}
\affiliation{Nuclear Physics Division, Bhabha Atomic Research Centre, Mumbai 400085, India.}
\author{G.~Mishra}
\affiliation{Nuclear Physics Division, Bhabha Atomic Research Centre, Mumbai 400085, India.}
\author{Debasish~Mondal}
\affiliation{Variable Energy Cyclotron Centre, 1/AF Bidhannagar, Kolkata 700064, India.}
\author{S.~Pal}
\affiliation{Pelletron Linac Facility, Tata Institute of Fundamental Research, Mumbai  400005, India.}
\author{M.S. Pose}
\affiliation{Department of Nuclear and Atomic Physics, Tata Institute of Fundamental Research, Mumbai 400005, India.}
\author{P.C.~Rout}
\affiliation{Nuclear Physics Division, Bhabha Atomic Research Centre, Mumbai 400085, India.}
\date{\today}

\begin{abstract}

Exclusive measurements of high energy $\gamma$-rays are performed in $\rm ^{124}Ba$ and $\rm ^{136}Ba$ at the same excitation energy ($\sim$49~MeV), to study properties of the giant dipole resonance (GDR) over a wider $N/Z$ range. The high energy $\gamma$-rays are measured in coincidence with the  multiplicity of low energy $\gamma$-rays  to disentangle the effect of temperature ($T$) and angular momentum ($J$). The GDR parameters are extracted employing a simulated Monte Carlo statistical model analysis. The observed $\gamma$-ray spectra of $\rm ^{124}Ba$ can be explained with prolate deformation, whereas a single component Lorentzian function which corresponds to a spherical shape could explain the $\gamma$-ray spectra from $\rm ^{136}Ba$. The observed GDR width in $\rm ^{136}Ba$ is narrower compared to that of $\rm ^{124}Ba$.
 The statistical model best fit GDR cross sections are found to be in  good agreement with the thermal shape fluctuation model (TSFM) calculations. Further, it is shown that the variation of GDR width with $T$ is well reproduced by the TSFM  calculations over the temperature range of 1.1--1.7~MeV.

\end{abstract}



\maketitle
\section{Introduction}

The isovector giant dipole resonance (GDR) is an excellent probe to study the collective behavior of the nucleus~\cite{snover, gaardhoje, schiller, drc0, berman, speth}. Using the GDR built on excited states (hot GDR) (produced via fusion-evaporation/inelastic scattering reactions/fission), the evolution of nuclear shape and damping mechanisms have been studied over a range of  excitation energy ($E^*$) and angular momentum ($J$) across the nuclear chart. Recent studies in $\rm ^{124,130,132}Sn$ nuclei ($N/Z$~=~1.48--1.64) showed significant difference in the dipole ($E\rm 1$) strength distribution. The occurrence of pygmy dipole resonance (PDR),  a resonance-like concentration of $E\rm 1$ strength above the neutron threshold energy has been confirmed in exotic nuclei $\rm ^{130,132}Sn$~\cite{adrich}. It would be interesting to see the effect of $N/Z$ asymmetry on $E$1 strength distribution in other nuclei having a large variation of $N/Z$ ratio.
 The barium isotopic chain ($A$~=~120-144) having a wide $N/Z$ ratio (1.14 to 1.57) and significant variation of ground state deformation ($\beta_{\rm gs}$~=~0.09--0.35~\cite{nndc}), provides an opportunity to study the GDR over a large isospin asymmetry. 
Earlier Vojtech $et$ $al.$~\cite{vojtech} have reported measurement of inclusive $\gamma$-ray spectra
in the decay of $\rm ^{124}Ba$ and $\rm ^{136}Ba$ nuclei produced using $\rm ^{12}C+^{112}Sn$ and $\rm ^{12}C+^{124}Sn$ reactions, respectively, at beam energies of 7.5 and 10.5~MeV/nucleon.
They have observed a single component GDR indicative of spherical shape for both nuclei with a large width of $\sim$~8 MeV, even though  these nuclei are known to be deformed in the ground state with $\beta_{\rm gs}$~=~0.301 and 0.1239 for $\rm ^{124}Ba$ and $\rm ^{136}Ba$, respectively~\cite{nndc}.   Further, the significant enhancement was observed in  the $\gamma$-ray yield beyond 20~MeV from $\rm ^{136}Ba$ nucleus, which was speculated to originate due to neutron skin in $\rm ^{136}Ba$. 
 It should be noted that in $A \sim$~130 region, a few measurements are reported addressing the saturation of the GDR width at high excitation energies in $\rm ^{132}Ce$~\cite{wieland} and in $\rm ^{136}Xe$~\cite{enders}. However, at these excitation energies the observed GDR widths have additional contributions from factors like compound nuclear lifetime. The exclusive data at lower excitation energies is desirable for $A \sim$~130 nuclei, for comparison with thermal shape fluctuation model (TSFM) predictions which is found to be most successful in describing the temperature dependence of GDR width in excited nuclei~\cite {alhassid, alhassidNPA1990, alhassidNPA1999, arumugam, gallardo}.

With this motivation, $\rm ^{12}C+^{112}Sn$ (at $E(\rm ^{12}C)$~=~64 MeV) and $\rm ^{12}C+^{124}Sn$ (at $E(\rm ^{12}C)$~=~52 MeV) reactions are performed to study the GDR in $\rm ^{124}Ba$ and $\rm ^{136}Ba$ nuclei, respectively, at the same excitation energy $\sim$~49~MeV. The choice of lower  excitation energy ensured that contributions from pre-equilibrium emission and nucleon-nucleon bremsstrahlung are expected to be negligible and will not affect the GDR spectra, enabling a cleaner comparison with TSFM.
The high energy $\gamma$-rays are detected in coincidence with low energy $\gamma$-rays for decoupling the temperature and angular momentum effects on the GDR parameters. The experimental GDR strength functions are compared with the GDR strength functions calculated using a thermal shape fluctuation model~\cite{alhassid, alhassidNPA1990, alhassidNPA1999, arumugam, gallardo} where the angular momentum and temperature dependence of shell effects are taken care. The GDR widths from the present measurement are combined with those from the earlier measurements~\cite{vojtech} to see the GDR width variation in wide $T$ range. The paper is organized as follows: Section II describes the details of the experiment and simulations of the detector response using GEANT4 tool-kit, followed by the statistical model analysis for extracting the GDR parameters in Section III.  The details of the TSFM calculation are presented in Section IV. In Section V, results of the GDR strength function  and comparison with the TSFM calculations are discussed. Finally, the summary and conclusion is presented in Section VI. 

\section{Experimental Details}
Pulsed $\rm ^{12}C$ beam of energies 64~MeV and 52~MeV from the Pelletron Linac Facility (PLF), Mumbai were used for GDR studies in $\rm ^{124}Ba$ and $\rm ^{136}Ba$ nuclei at the same excitation energy ($\sim$~49~MeV) employing self-supporting $\rm ^{112}Sn$ (2.3~mg/cm$\rm ^2$) and $\rm ^{124}Sn$ (1.9~mg/cm$\rm ^2$) targets, respectively. The detector system used for high energy $\gamma$-ray measurement is similar to that described in Ref.~\cite{cghosh}. High energy $\gamma$-rays were detected in an array of seven close-packed hexagonal BaF$\rm _2$ detectors (each having face-to-face distance of 9~cm and length of 20~cm) placed at distance of $\sim$~57~cm from the target and at an angle of 125$^\circ$  with respect to the beam direction. The array was surrounded with an annular plastic scintillator for cosmic ray rejection. The entire detector setup (BaF$\rm _2$+plastic) was surrounded by 10~cm thick lead for attenuating ambient and beam induced $\gamma$-ray background. In addition, a 5~mm thick lead sheet was mounted on the front face of BaF$\rm _2$ array for reducing low-energy $\gamma$-ray and x-ray background. The beam dump was $\sim$~2~m away from the target and was properly shielded with borated paraffin and lead for reducing the neutron and $\gamma$-ray background, respectively. The time of flight (TOF) of each BaF$\rm _2$ detector with respect to the RF pulse was used to separate neutron and $\gamma$-ray induced events. The typical full width at half maximum of the $\gamma$-ray prompt peak in the TOF spectrum is $\sim$~2~ns. Each detector anode signal was integrated over two different gates: 2~$\mu$s referred to as $Q_{\rm long}$ and 200~ns referred to as $Q_{\rm short}$ for energy measurement and pileup rejection using pulse shape discrimination, respectively. The array was calibrated using $\rm ^{137}Cs$, $\rm ^{60}Co$ and $\rm ^{241}Am$-$\rm ^9Be$ radioactive sources covering energy range 0.6--4.4~MeV. The higher energy calibration points (18.1 and 22.6~MeV) were obtained by bombarding a $\sim$~1~mg/$\rm cm^2$ thick $\rm ^{11}B$ target (prepared by electro-deposition on a 127 $\mu$m thick tantalum backing) by a proton beam of energy 7.2~MeV~\cite{cecil}. During the  experiment, the gain of the BaF$\rm _2$ detectors were periodically monitored using radioactive sources and the variation was found to be within $\pm$ 2\%.

The energy response of the array was generated using GEANT4~\cite{geant} based simulation employing the actual detector configuration in the present experiment. The simulated energy spectra of each detector were convoluted with Gaussian resolution function for comparison with the measured $\gamma$-ray spectra from sources and $\rm ^{11}B(p,\gamma)$ reaction, where the FWHM of the Gaussian function was optimized to fit the experimental spectra. The resolution (FWHM/$E$) was found to vary as 1/$\sqrt{E}$ with a typical value of $\sim$~6\% at 22~MeV.
For a given incident energy, the individual detector spectra folded with the resolution function were added to generate the total energy spectrum and the response matrix of the array was constructed for a range of $\gamma$-ray energies $E_{\gamma}$~=~1 to 30 MeV.

In the present measurement, the angular momentum of the compound nucleus (CN) is extracted from the multiplicity of low energy $\gamma$-rays. An array of 14 element hexagonal BGO detectors (each having 5.6~cm face-to-face distance and length of 7.6~cm), arranged in castle geometry above and below the target chamber (7+7), was used as the multiplicity filter. Fold, number of BGO detectors fired above 120~keV threshold within 50~ns coincidence window, is a measure of the multiplicity and is recorded for each event. The logic `OR' of the (time-matched) timing signals from top and bottom BGOs  is used for TOF ($BGO\_TOF$) measurement with respect to the RF-pulse. This is also used for subtracting the chance coincident events in the multiplicity filter.  The multiplicity to fold response of the array depends on the efficiency and the cross-talk of the array. The efficiency of the array was measured using a calibrated $\rm ^{137}Cs$ source placed at the target position and found to be 64.2$\pm$0.2\%. The cross-talk was obtained in a coincidence measurement using $\rm ^{60}Co$ source placed at the target position and a CeBr$\rm_3$ detector (cylindrical--38~mm$\times$38~mm) placed outside of the target chamber. The measured cross-talk probability at this energy is $\sim$~9\%. The response of the multiplicity filter was also calculated by GEANT4 based Monte Carlo simulations. The simulated efficiency (63\%) at 662~keV and cross-talk probability (8\%) at 1.2~MeV are in good agreement with measurements.

The master trigger was generated when the sum energy deposited in the BaF$\rm _2$ array was above $\sim$~5~MeV. Parameters recorded for each event were $Q_{\rm short}$, $Q_{\rm long}$, TOF of each BaF$\rm_2$ (${BaF\_TOF}$) detector, $BGO\_TOF$ and fold ($F$). Event-by-event data were acquired using a CAMAC-based acquisition-cum-analysis software LAMPS~\cite{lamps} for 0.23 pmC and 0.63 pmC of incident beam particles in $\rm ^{12}C+^{112}Sn$ and $\rm ^{12}C+ ^{124}Sn$ reactions, respectively.

In the offline analysis, the $Q_{\rm long}$ for individual BaF$\rm _2$ detector was filtered using the $\gamma$-prompt in  ${BaF\_TOF}$ and pileup rejection condition.  The fold was also filtered using the prompt in $BGO\_TOF$. Both the filtered $Q_{\rm long}$ and fold were corrected for the chance coincidences in ${BaF\_TOF}$ and $BGO\_TOF$, respectively. The total energy spectrum was constructed by adding these corrected $Q_{\rm long}$ of individual detector after calibration. The spectrum is further corrected for the Doppler effect due to finite recoil velocity of the CN corresponding to a mean angle of 125$^\circ$ for average CN recoil velocity ($v/c \sim$~0.01).
\begin{figure}[!ht]
\includegraphics[scale=0.34]{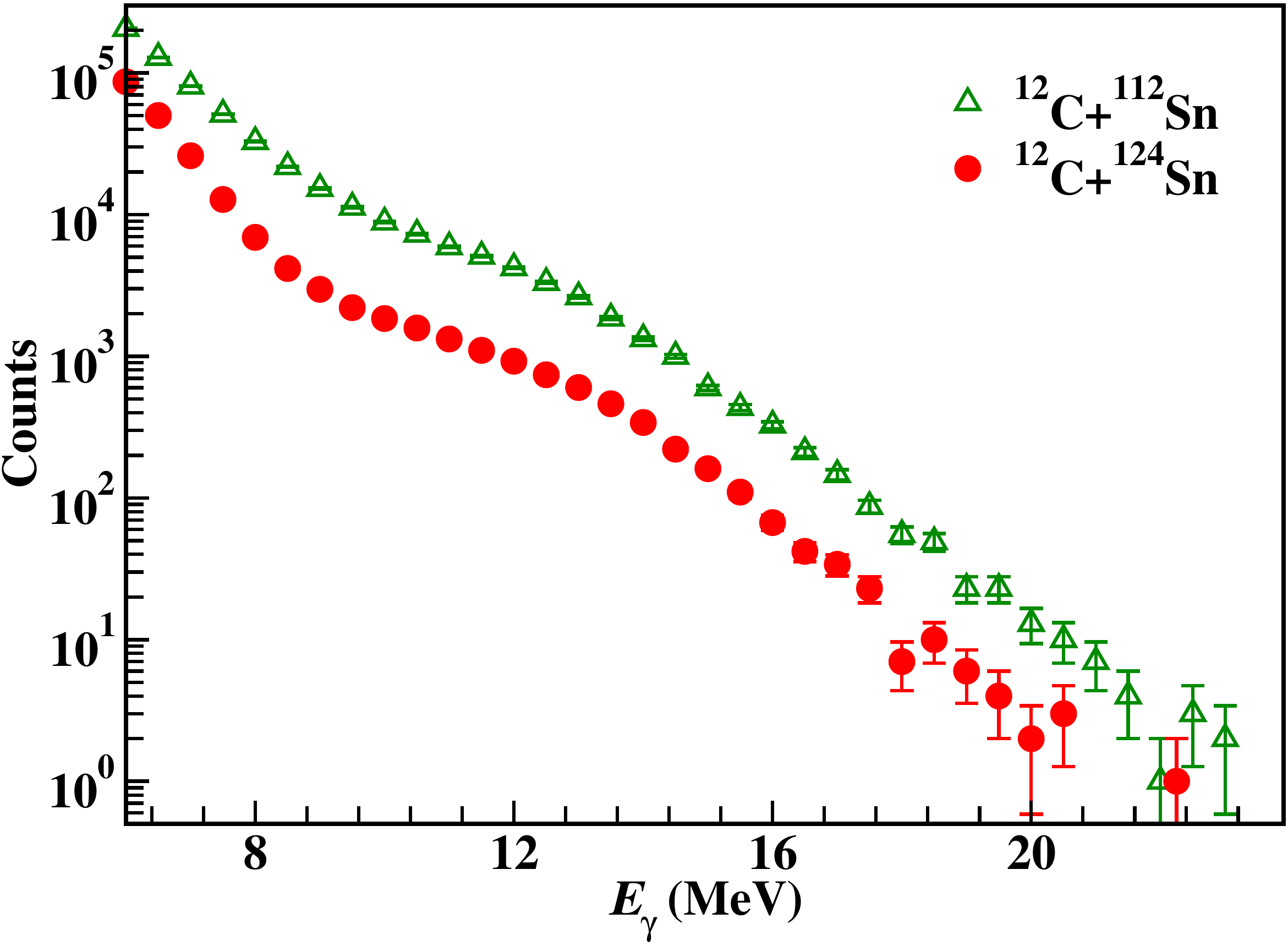}
\caption{\label{logplot}(Color online) The high energy $\gamma$-ray spectra from  $\rm ^{12}C+ ^{112}Sn$ reaction at 64~MeV (triangle) and from  $\rm ^{12}C+ ^{124}Sn$ reaction at 52~MeV (solid circle), for the fold window 5-6.}
\end{figure}
 The $\gamma$-ray spectra  from $\rm ^{12}C+ ^{112}Sn$ reaction at 64~MeV and from $\rm ^{12}C+ ^{124}Sn$ reaction at 52~MeV, for the fold window 5-6, are shown in Fig.~\ref{logplot}. From the events satisfying the present analysis conditions, a two dimensional matrix of fold vs total energy was generated. Suitable projection of the matrix yielded desired fold-gated $\gamma$-ray spectrum and energy-gated fold distribution.

The larger fusion cross section of the beam with light mass impurities present in the target (mainly C and O) can give significant contribution in the high energy $\gamma$-ray spectrum. The amount of $\rm ^{12}C$ and $\rm ^{16}O$ present in the target was estimated from the yield of 4.44 MeV and 6.13~MeV $\gamma$-rays in the resonance radiative proton capture reactions at $E_{\rm p}$~=~7.78~MeV~\cite{barnard} and at $E_{\rm p}$~=~7.46~MeV~\cite{dangle}, respectively. The above reactions with $\rm ^{12}C$ and WO$\rm _3$ target were used as reference for 4.4 and 6.13~MeV $\gamma$-ray yield, respectively. From the ratios of the $\gamma$-ray yields from the reactions using actual target, and $\rm ^{12}C$ and WO$_{\rm 3}$ targets, the carbon and oxygen impurities in the target were extracted and were found to be $\sim$~4\% and $\sim$~14\%, respectively. However, the contribution from these impurities at fold gated ($F~>$~3) high energy $\gamma$-ray spectrum was less than 1\% and hence was ignored.
\section{Statistical Model Analysis}
\begin{figure}
\includegraphics[scale=0.34]{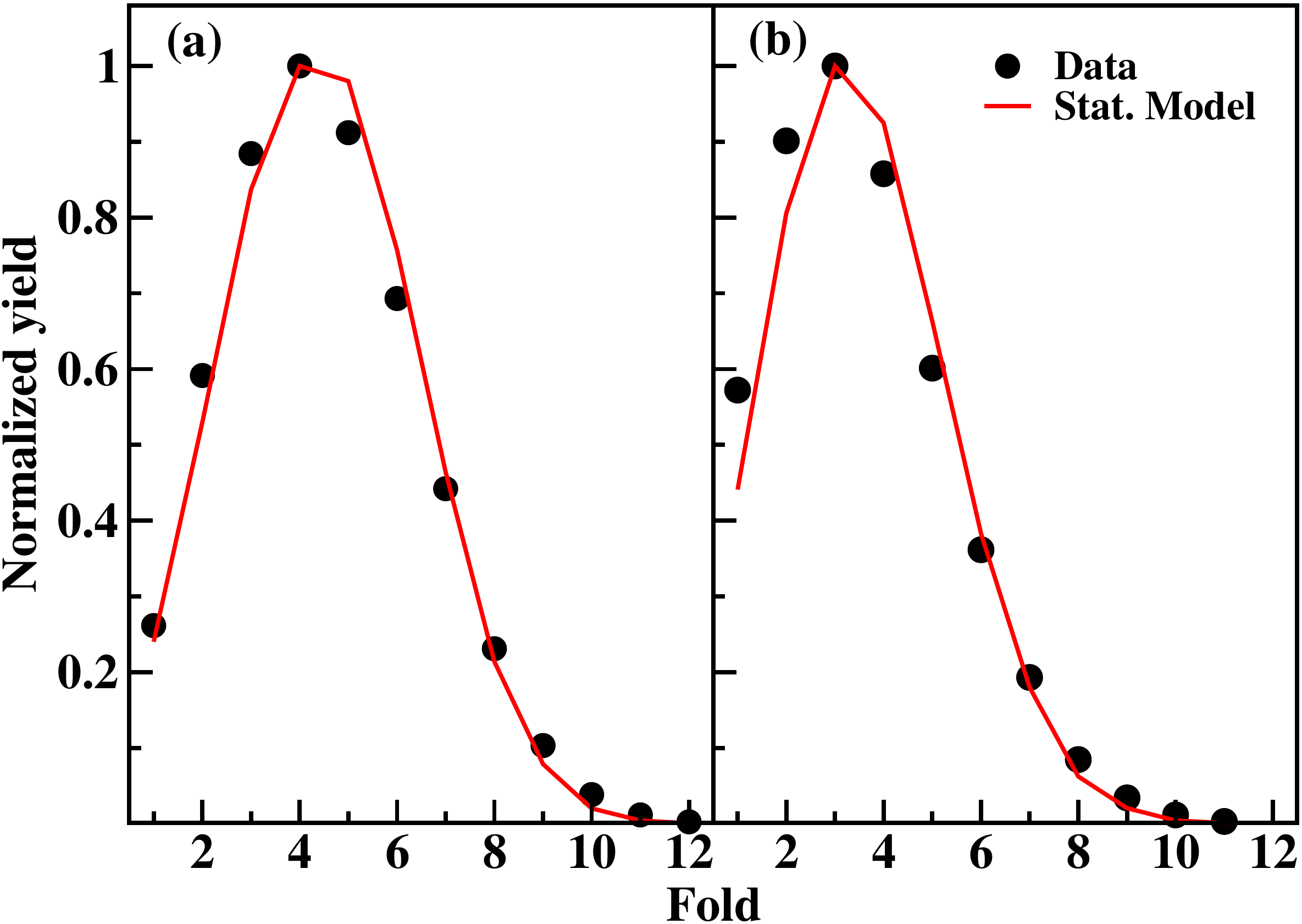}
\caption{\label{fold_dist}(Color online) The energy gated (10-20~MeV) experimental and simulated fold distribution for reaction: (a) $\rm ^{12}C+^{112}Sn$ with $E(\rm ^{12}C)$~=~64~MeV  and (b) $\rm ^{12}C+ ^{124}Sn$ with $E(\rm ^{12}C)$~=~52~MeV.}
\end{figure}

The statistical model (SM) analysis is carried out for extracting the GDR parameters from the experimental fold-gated high energy $\gamma$-ray spectrum. For generating the fold-gated $\gamma$-ray spectrum, a simulated Monte Carlo CASCADE (SMCC)~\cite{drc3} code is used. The parameters used for calculating the particle transmission coefficients in optical model potential are taken from Refs.~\cite{wilmore,perey,mcfadden}. Another important input for the SM calculation is the nuclear level density. For the present work, the level density formalism proposed by Igantyuk $\it et$ $\it al.$~\cite{ignatyuk} is used with asymptotic level density parameter  $\it \tilde{a}$~=~$\it A$/9.0~$\rm MeV^{-1}$ for both $\rm ^{124,136}Ba$ nuclei. In these calculations, the CN is assumed to follow the standard angular momentum distribution:
\begin{equation}
\sigma(J_{\rm CN}) = \sigma_{0} \frac{2J_{\rm CN}+1}{1+exp[(J_{\rm CN}-J_0)/\delta J]}\hspace{2pt},
\end{equation}
with $\delta J$~=~2. The residue spin distribution ($J_{\rm res}$) is calculated for each $J_{\rm CN}$ as a function of $\gamma$-ray energy by summing over all intermediate $\gamma$-decays. The multiplicity ($M$) of low energy $\gamma$-rays for the decay from spin $J_{\rm res}$ to the ground state is calculated using relative decay probability ($P_{\rm r}$) of $\Delta J$~=~1 and $\Delta J$~=~2 transitions as a parameters. The multiplicity to fold response of the BGO array is constructed in a Monte Carlo approach incorporating the energy dependent efficiency and cross-talk probability calculated using GEANT4 simulation tool-kit. By varying $P_{\rm r}$, the experimental fold distributions in $\rm ^{12}C+^{112}Sn$ and $\rm ^{12}C+^{124}Sn$ reactions are reproduced and are shown in Fig.~\ref{fold_dist}.

The best fit GDR parameters for describing the different fold gated $\gamma$-ray spectra are extracted following the same procedure as described in Ref.~\cite{cghosh}.
The photo-absorption cross section for an axially symmetric deformed (prolate or oblate) nucleus can be expressed by sum of two-component Lorentzian function as
\begin{equation}
\label{sigmaabs}
\sigma_{\rm abs}(E_\gamma) = \frac{4\pi e^2\hbar}{mc}\frac{NZ}{A}\sum\limits_{j=1}^{j=2} \frac{S_j\it \Gamma_jE_\gamma^2}{(E_\gamma^2-E_{j}^2)^2+E_\gamma^2\it \Gamma_j^2}\hspace{2pt},
\end{equation}
where $\it N$ and $\it Z$ represent the neutron and proton numbers, $E_{\rm 1(2)}$, $\it \Gamma_{\rm 1(2)}$ and $S_{\rm 1(2)}$ are the centroids, widths and strengths for the two components, respectively.  For spherical nuclei, the oscillations along three mutually perpendicular axes are identical and single component Lorentzian in Eq.~\ref{sigmaabs} can describe the observed GDR strength distribution.  
It is assumed that the GDR oscillation exhausts 100\% ($S_{\rm 1}+S_{\rm 2}=1$) of the Thomas-Reiche-Kuhn (TRK) sum rule~\cite{trk}. It may be mentioned that in earlier work we have shown that the extracted best fit parameters remain unaffected within errors even with  25\% variation of the TRK sum rule~\cite{cghosh}.
The experimental fold gated $\gamma$-ray spectra are compared with the calculated spectra after folding the BaF$\rm _2$ response function. Since the absolute $\gamma$-ray cross sections are not measured, both the spectra are normalized at $E_{\rm \gamma}$~=~8~MeV. The $\chi^2$ minimization and visual inspection in the energy range of 8--23~MeV are used to achieve the goodness of fit and to extract the best fit GDR parameters.

 From the projectile energy and the $Q$-value of the reaction the excitation energy of the compound nucleus is calculated. If $E_{\rm rot}$ and $\Delta_{\rm p}$ are the rotational energy and pairing energy, respectively, then the net excitation energy available for internal excitation is $\it U=E^*_{\rm f}-E_{\rm rot}-\Delta_{\rm P}$,  where $\it E^*_{\rm f}$ is the excitation energy after the emission of the GDR $\gamma$-ray. The temperature of the state on which the GDR built is calculated using the relation $\it U = aT^2$, where $\it a(U)$ is the Ignatyuk level density parameter~\cite{ignatyuk}. The average temperature ($<T>$) and angular momentum ($<J>$) for different fold windows are calculated following the same procedure as in Ref.~\cite{cghosh}.

\begin{figure}
\vspace{5pt}
\includegraphics[scale=0.34]{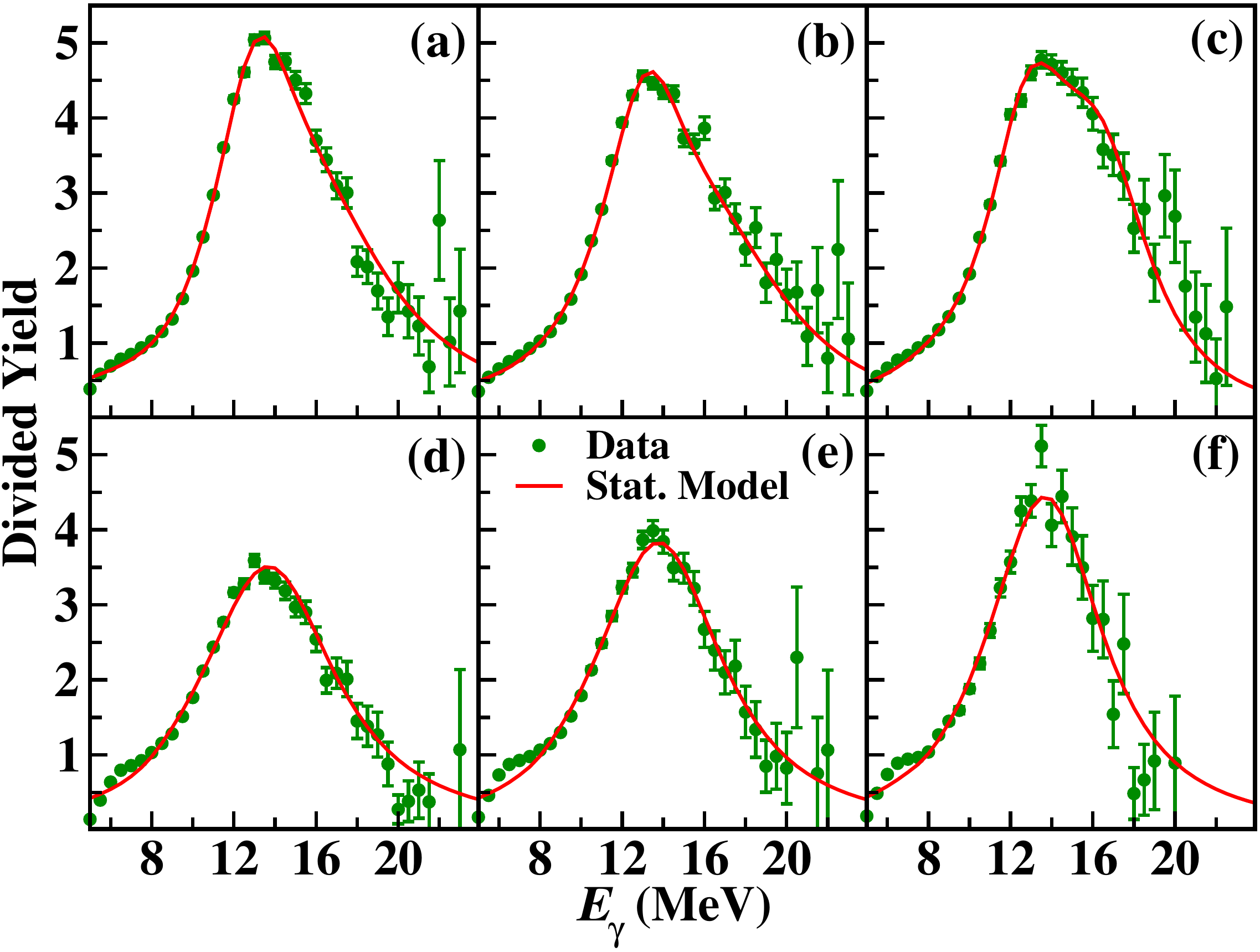}
\caption{\label{dividedplot}(Color online) 
Top panels show the divided plots for $\rm ^{12}C+ ^{112}Sn$ - (a)fold 3-4,  (b) 5-6  and (c) 7-14;   Bottom panels (d),(e), (f) show the same  for $\rm ^{12}C+ ^{124}Sn$. The line corresponds to the statistical model best fit calculations.}
\end{figure}

 For the present statistical model analysis, the $\gamma$-ray spectra are calculated  considering prolate, oblate and spherical shapes.
 It is observed that while in the  earlier work \cite{vojtech} the high energy $\gamma$-ray spectra from $\rm ^{124}Ba$  were described with  spherical shape in statistical model, the present data for $\rm ^{124}Ba$ could not be fitted with a single component Lorentzian function.  
  The $\gamma$-ray spectra of $\rm ^{124}Ba$ are found to be consistent with prolate deformation, whereas  for $\rm ^{136}Ba$ 
a single component Lorentzian function corresponding to  spherical shape could explain the  data.
The $\gamma$-ray spectrum calculated with an arbitrary constant dipole strength of 0.2~W.u, folded  with the BaF$\rm _2$ response, was used for generating divided plot. 
The divided plots for both the experimental and the calculated $\gamma$-ray spectra corresponding to different fold windows  are shown in Fig.~\ref{dividedplot}. 

 The best fit parameters for $\rm ^{124}Ba$ are listed in Table~\ref{tab:gdrpara124ba}.
 In case of axially symmetric deformed nucleus, the centroid of GDR ($E_{\rm D}$)  is calculated  as the weighted average of the centroid of two components and the effective GDR width ($\it \Gamma_{\rm D}$) is taken as the full width at half maximum of the total GDR strength function. In Table~\ref{tab:gdrpara124ba1}, the $E_{\rm D}$, $\it \Gamma_{\rm D}$, $<T>$, $<J>$ and deformation parameter ($\beta$) for $\rm ^{124}Ba$ are tabulated for different fold windows. For $\rm ^{124}Ba$ nucleus, within the limited $T$ and $J$ range studied in present experiment, the GDR centroid ($\sim$~16~MeV), width ($\sim$~8.0~MeV) and $\beta$ remain constant.  
 
\begin{table}[!ht]
\caption{\label{tab:gdrpara124ba} Best fit GDR parameters from SMCC analysis with prolate deformation for various fold windows for $\rm ^{124}Ba$ at $\sim$~49~MeV excitation energy.}
\begin{ruledtabular}
\begin{tabular}{cccccc}
Fold & $E_1 (\rm MeV)$ & $\it \Gamma_1 (\rm MeV)$&
 $E_2 (\rm MeV)$  &$\it \Gamma_2 (\rm MeV)$ & $S_2$\\ \hline
3-4& 13.6(1)& 3.7(2)& 16.7(2)  &9.0(4)& 0.70(3)\\
5-6& 13.7(1)& 4.0(2)& 17.5(3)&9.5(5)& 0.67(3)\\
7-14& 13.4(1)& 3.7(2)& 17.1(2)&5.9(3)& 0.70(3)\\
\end{tabular}
\end{ruledtabular}
\end{table}

\begin{table}[!ht]
\caption{\label{tab:gdrpara124ba1}Extracted GDR parameters and nuclear deformation $\beta$, as a function of $\it J$ and $\it T$ for prolate deformation for $\rm ^{124}Ba$.}
\begin{ruledtabular}
\begin{tabular}{cccccc}
Fold & $<J>$ & $<T>$& $E_{\rm D}$  &$\it \Gamma_{\rm D}$ &$\beta$\\
 &($\rm \hbar$) & $ (\rm MeV)$ & $ (\rm MeV)$ & $ (\rm MeV)$ & \\ \hline
3-4& 14(7) & 1.19(27)& 15.8(2)&7.9(4) &0.24(2)\\
5-6& 19(6)& 1.16(25) & 16.2(3)&8.8(5) &0.29(2) \\
7-14& 22(6)& 1.12(24) & 16.0(2)&8.2(4) &0.29(2)\\
\end{tabular}
\end{ruledtabular}
\end{table}

For $\rm ^{136}Ba$, the extracted best fit parameters for various fold windows are listed in Table~\ref{tab:gdrpara136ba} along with $<T>$ and $<J>$.  In this case also, the centroids ($\sim$~14.7~MeV) and the width are nearly constant in the present  $T$ and $J$ range and  width  is significantly narrower (by $\sim$~1~MeV) than that for $\rm ^{124}Ba$. The observed centroid ($\sim$~16~MeV) for $\rm ^{124}Ba$ is found to be reasonable agreement with the ground state systematics, whereas that for $\rm ^{136}Ba$ ($\sim$~14.7~MeV) is marginally smaller than systematics.
\begin{table}[h]
\caption{\label{tab:gdrpara136ba} Best fit GDR parameters from SMCC analysis with prolate deformation for various fold windows for $\rm ^{136}Ba$ at $\sim$~49~MeV excitation energy.}
\begin{ruledtabular}
\begin{tabular}{ccccc}
Fold & $<J>$ & $<T>$& $E_{\rm D}$  &$\it \Gamma_{\rm D}$ \\
 &($\rm \hbar$) & $ (\rm MeV)$ & $ (\rm MeV)$ & $ (\rm MeV)$\\ \hline
3-4&11(5)& 1.24(28)& 14.7(2)& 7.4(3)\\
5-6&13(5)&1.23(28)& 14.7(2)& 7.0(3)\\
7-14&15(5)&1.22(29)& 14.5(2)& 6.7(3)\\
\end{tabular}
\end{ruledtabular}
\end{table}

\section{TSFM calculation}

\begin{figure}[h]
\includegraphics[width=.3\columnwidth, clip=true]{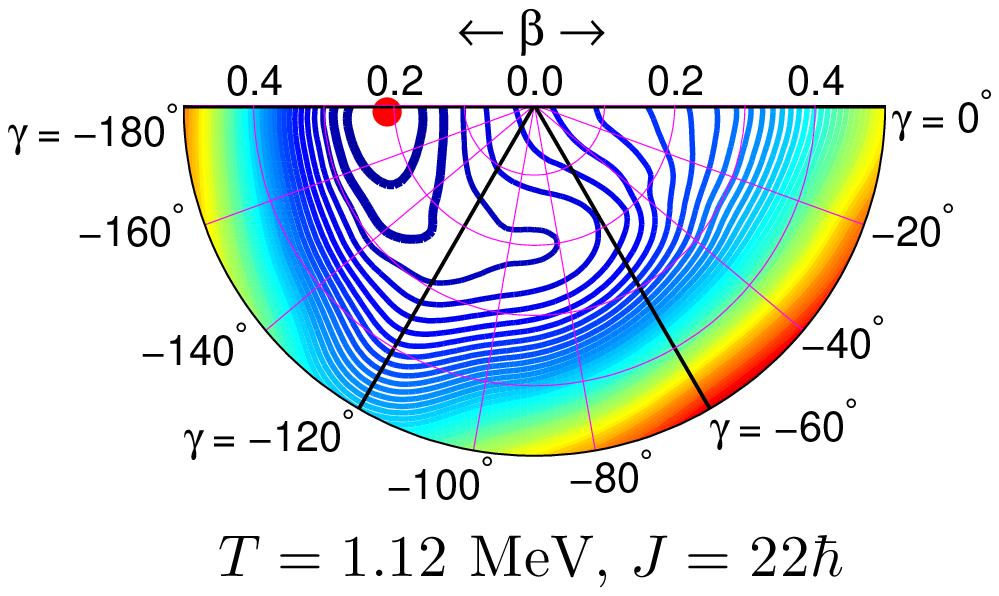}
\includegraphics[width=.3\columnwidth, clip=true]{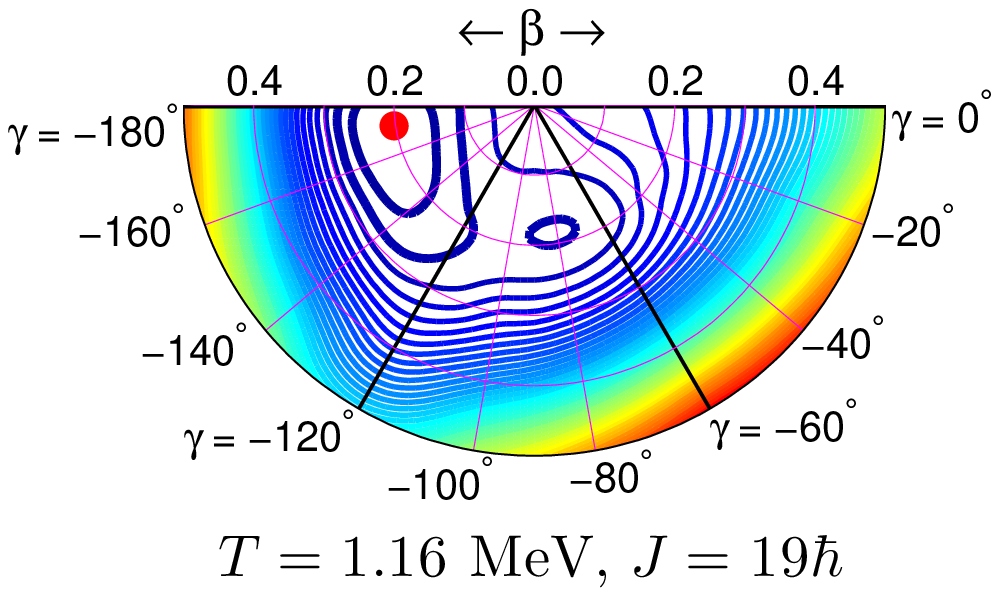}
\includegraphics[width=.3\columnwidth, clip=true]{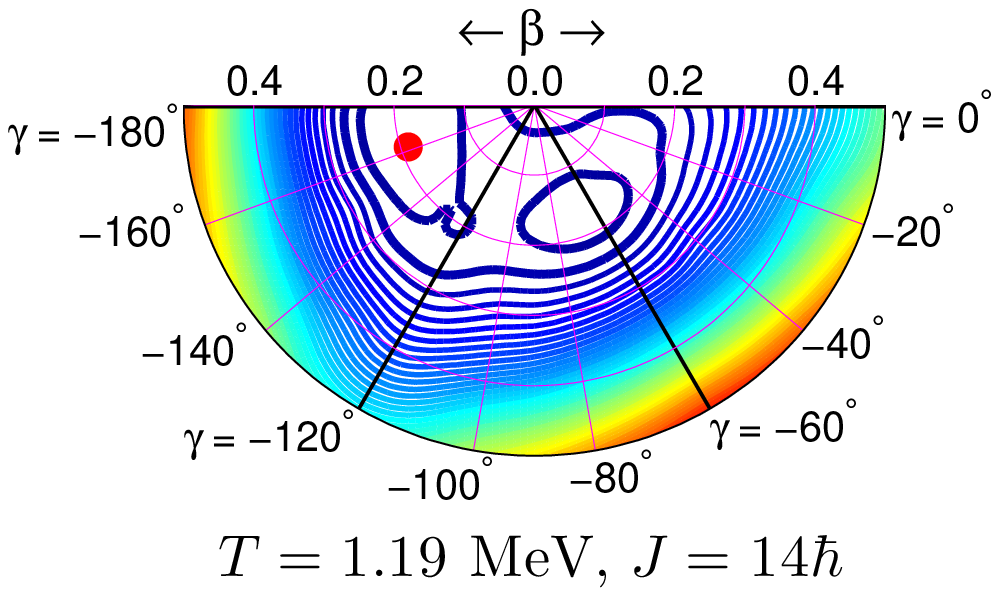}
 \vskip-26pt
 \hbox{\hspace{20pt}(a) \hspace{125pt}(b)\hspace{130pt}(c)}
 \vskip10pt
\caption{(Color online) The free energy surfaces (FES) of $^{124}$Ba at different temperature ($T$) and angular momentum ($I$) combinations corresponding
to the data measured at beam energy $E\sim$~64~MeV.
In this convention, $\gamma=0^\circ$ and $-120^\circ$ represent the non-collective and collective prolate shapes, respectively; $\gamma=-180^\circ$ and $-60^\circ$ represent the non-collective and collective oblate shapes, respectively. The contour line spacing is 0.2 MeV. The most probable shape is represented by a filled circle and first two minima are represented by thick lines.}
\label{Ba124}
\end{figure}

\begin{figure}[tbp]
\includegraphics[width=.3\columnwidth, clip=true]{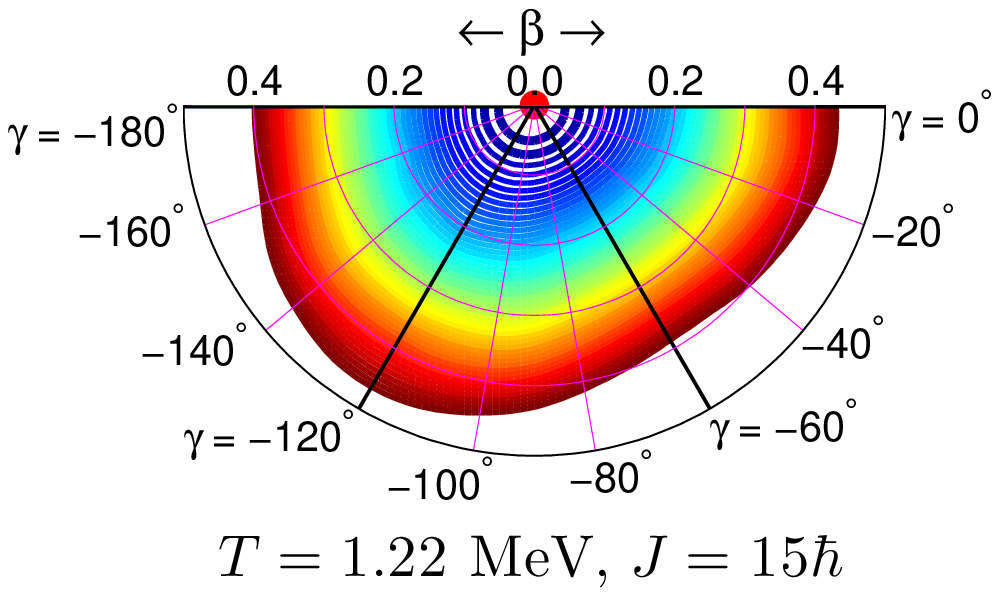}
\includegraphics[width=.3\columnwidth, clip=true]{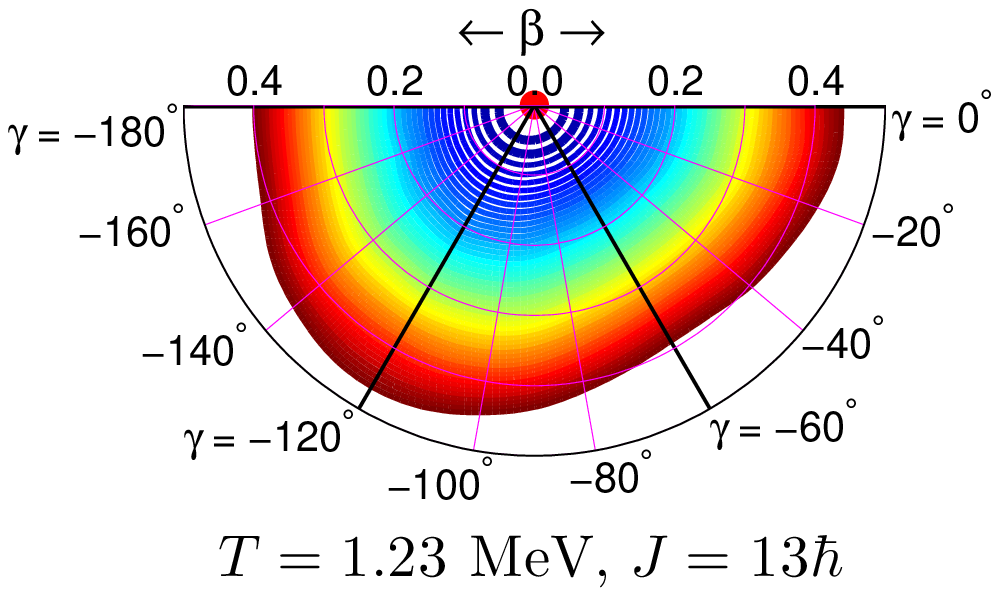}
\includegraphics[width=.3\columnwidth, clip=true]{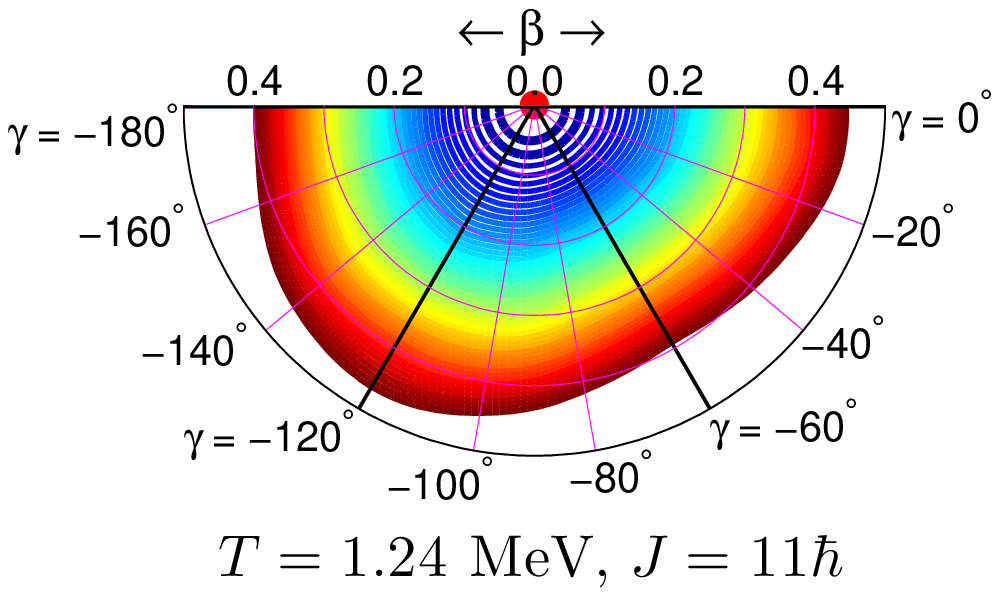}
 \vskip-26pt
 \hbox{\hspace{20pt}(a) \hspace{125pt}(b)\hspace{130pt}(c)}
\vskip10pt
\caption{(Color online) Same as Fig.~\ref{Ba124}, but for $^{136}$Ba.}
\label{Ba136}
\end{figure}

The free energy surfaces (FES) for $^{124}$Ba and $^{136}$Ba nuclei are calculated at the experimentally measured values of $T$ and $J$ and are shown in Figs.~\ref{Ba124} and~\ref{Ba136}, respectively.
The most probable shape (MPS) of the nucleus is represented by a red solid dot. In the case of $^{124}$Ba, at $T=1.12$ MeV and $J=22\hbar$ the nucleus prefers an oblate deformation with $\beta=0.2$ and $\gamma=-180^\circ$. As $T$ increases, our calculations suggest a shallow bottom with  multiple minima in the FES. At $T=1.16$ MeV and $J=19\hbar$, the nucleus prefers an oblate shape with $\beta=0.2$ and $\gamma=-170^\circ$ with an additional minimum in the FES at $\gamma=-80^\circ$. At $T=1.19$ MeV and $J=14\hbar$, the nucleus acquires a near triaxial shape with $\beta=0.2$ and $\gamma=-160^\circ$ with a coexisting minimum in the FES. The deformation parameter $\beta$ remains constant at all measured $T$ values. At higher $T$ values, the area spanned by the first and second minimum contour lines are large. But the area covered by these contour lines lies in the lower $\beta$ regions. This leads to a decrease in the average deformation values even if there is an increase in the thermal shape fluctuations.       
For $^{136}$Ba nucleus the FES (shown in Fig.~\ref{Ba136}) are more crisp and the shape of the nucleus remains spherical in the measured range of $T$ and $J$.

The thermal fluctuations related to the shape degrees of freedom  at a finite excitation energy are large in nucleus. The thermal shape fluctuations (TSF) carry information about the shape rearrangements~\cite{alhassidNPA1990} at finite excitation energy. The general expression for the average value of GDR cross section, $\sigma$ incorporating such TSF has the form~\cite{alhassidNPA1991,alhassidNPA1999} 
\begin{equation}\label{average_1}
\langle \sigma(T,J)\rangle _{\beta ,\gamma }=\frac{\int_\beta\int_\gamma \mathcal{D}[\alpha
]\exp\left[{-F_\mathrm{TOT}(T, J;\beta ,\gamma )/T}\right]\Im_{\mathrm{TOT}}^{-3/2}\sigma(J;\beta,
\gamma )}{\int_\beta\int_\gamma \mathcal{D}[\alpha
]\exp\left[{-F_\mathrm{TOT}(T,J; \beta ,\gamma )/T}\right]\Im_{\mathrm{TOT}}^{-3/2}}\;,
\end{equation}%
with the volume element given by $\mathcal{D}[\alpha ]=\beta ^{4}|\sin 3\gamma |\,d\beta \,d\gamma$. $F_\mathrm{TOT}$ is the free energy calculated by Microscopic-Macroscopic method with proper $T$ and $J$ dependent shell corrections and $\Im_{\mathrm{TOT}}$ is the moment of inertia. 
 
We employ a macroscopic approach for GDR to relate the GDR observables and nuclear shapes \cite{arumugam,shanmugam,rhinePRC2015}. The GDR Hamiltonian
could be written as sum of the anisotropic harmonic oscillator Hamiltonian
and the separable dipole-dipole interaction,
\begin{equation}
H=H_{osc}+\eta D^{\dagger} D\;. \label{GDR hamiltonian1}
\end{equation}
The value of dipole-dipole interaction strength $\eta$ was varied to get the best fit to data ($E_{\gamma}$~=~8 to 23 MeV) and is chosen as $2.9$ for $^{124}$Ba and $2.8$ for $^{136}$Ba, respectively.

\section{Results and Discussion}
 The comparison of fold gated GDR spectra in $\rm ^{124}Ba$ and $\rm ^{136}Ba$, presented in earlier section, brings out significant differences between these two nuclei. The spherical shape of $\rm ^{136}Ba$ could be a manifestation of a nearly closed shell $N$~=~82 and is also  consistent with the FES calculations as can be seen in Fig.~\ref{Ba136}.
For $\rm ^{124}Ba$, the $\beta \sim$~0.29(2) obtained from the SM analysis is similar to  $\beta_{\rm gs}= 0.301$~\cite{nndc}. This is not surprising since the the yrast-like deformation is expected to persist upto  $T_{\rm lim} \sim 40 \delta A^{-1/3}$~\cite{bjornholm} (where $\delta \approx 0.95\beta$), which is $\sim$~2~MeV for $\rm ^{124}Ba$. However, the TSFM predicts the equilibrium deformation to be oblate. 
 
\begin{figure}[!ht]
\vspace{5pt}
\includegraphics[scale=0.34]{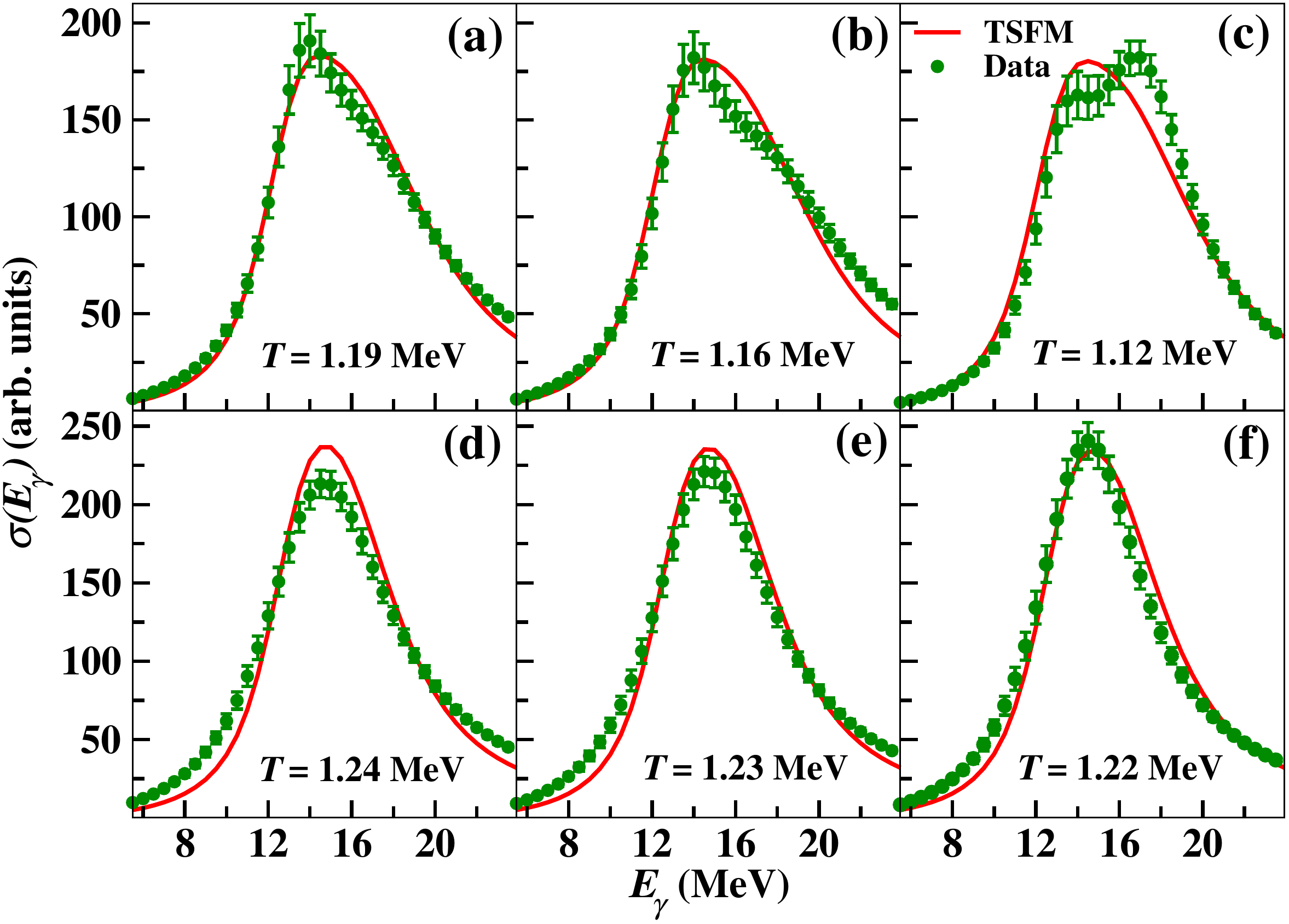}
\caption{\label{strengtplot}(Color online)  Comparison of $\sigma_{\rm stat}$ and $\sigma_{\rm TSFM}$ for fold windows 3-4, 5-6 and 7-14 for $\rm ^{12}C+^{112}Sn$ reaction at $E(\rm ^{12}C)$~=~64~MeV (in the top panels--a, b, c) and for $\rm ^{12}C+^{124}Sn$ reaction at $E(\rm ^{12}C)$~=~52~MeV (in the bottom panels--d, e, f).}
\end{figure}

For comparison of the data with TSFM, the procedure described in Ref.~\cite{cghosh} is followed, where the  experimental GDR cross section ($\sigma_{\rm stat}$) represented by the photo-absorption cross section used as input in the SM analysis (Eq.~2), is compared with the calculated GDR cross section using TSFM ($\sigma_{\rm TSFM}$). For each fold window, the $\sigma_{\rm stat}$ is calculated using the best fit parameters and is normalized to $\sigma_{\rm TSFM}$ in the $E_\gamma$~=~8--23~MeV range, since the absolute GDR cross section is not measured. 
The comparison of $\sigma_{\rm stat}$ and $\sigma_{\rm TSFM}$ 
 for $\rm ^{124}Ba$ and $\rm ^{136}Ba$ is shown in top and bottom panels of  Fig.~\ref{strengtplot}, respectively, for different fold windows. The error bars in $\sigma_{\rm stat}$ represent the variation of $\sigma_{\rm stat}$ calculated from the errors of best fit parameters.

 It is evident that for $\rm ^{136}Ba$  the agreement with TSFM is very good. For $\rm ^{124}Ba$, some differences are seen in the shape of strength function since the FES predicts the equilibrium shape as oblate (see Fig.~\ref{Ba124}). 
 However, the observed effective GDR width is in reasonable agreement with that obtained from the TSFM calculation. The $\it \Gamma_{\rm D}$ in $\rm ^{136}Ba$ is significantly narrower ($\sim$~15\%)  compared to that in $\rm ^{124}Ba$, emphasizing the dominant role of fluctuations due to shallow minima on FES (Fig.~\ref{Ba124}) in the latter case.

From earlier measurement by Vojtech $et$ $al.$~\cite{vojtech}, it was found that $\it \Gamma_{\rm D}$ in $\rm ^{124}Ba$ and $\rm ^{136}Ba$ are similar at higher excitation energy.
In order to compare the GDR widths from the inclusive measurements of earlier data~\cite{vojtech}, 
the average temperatures ($T$) after the GDR emission are calculated for the compound nuclei $\rm ^{124,136}Ba$ following the procedure described in Refs.~\cite{wieland1,srijit_PRC2008}. The net energy available for internal excitation for these nuclei after the GDR $\gamma$-ray emission is calculated by subtracting the energy loss due to pre-equilibrium particle emission~\cite{kelly_PRL1999} and the rotational energy. The rotational energies are calculated using the moment of inertia of a symmetrically deformed nuclei with the respective ground state deformation.
 It should be mentioned that we obtain a lower value of $T$ for $E(\rm ^{12}C)$~=~127 MeV, as compared to Fig.~3 in Ref.~\cite{wieland},  due to incorporation of the pre-equilibrium emission.
 
\begin{figure}[!ht]
\vspace{5pt}
\includegraphics[scale=0.34]{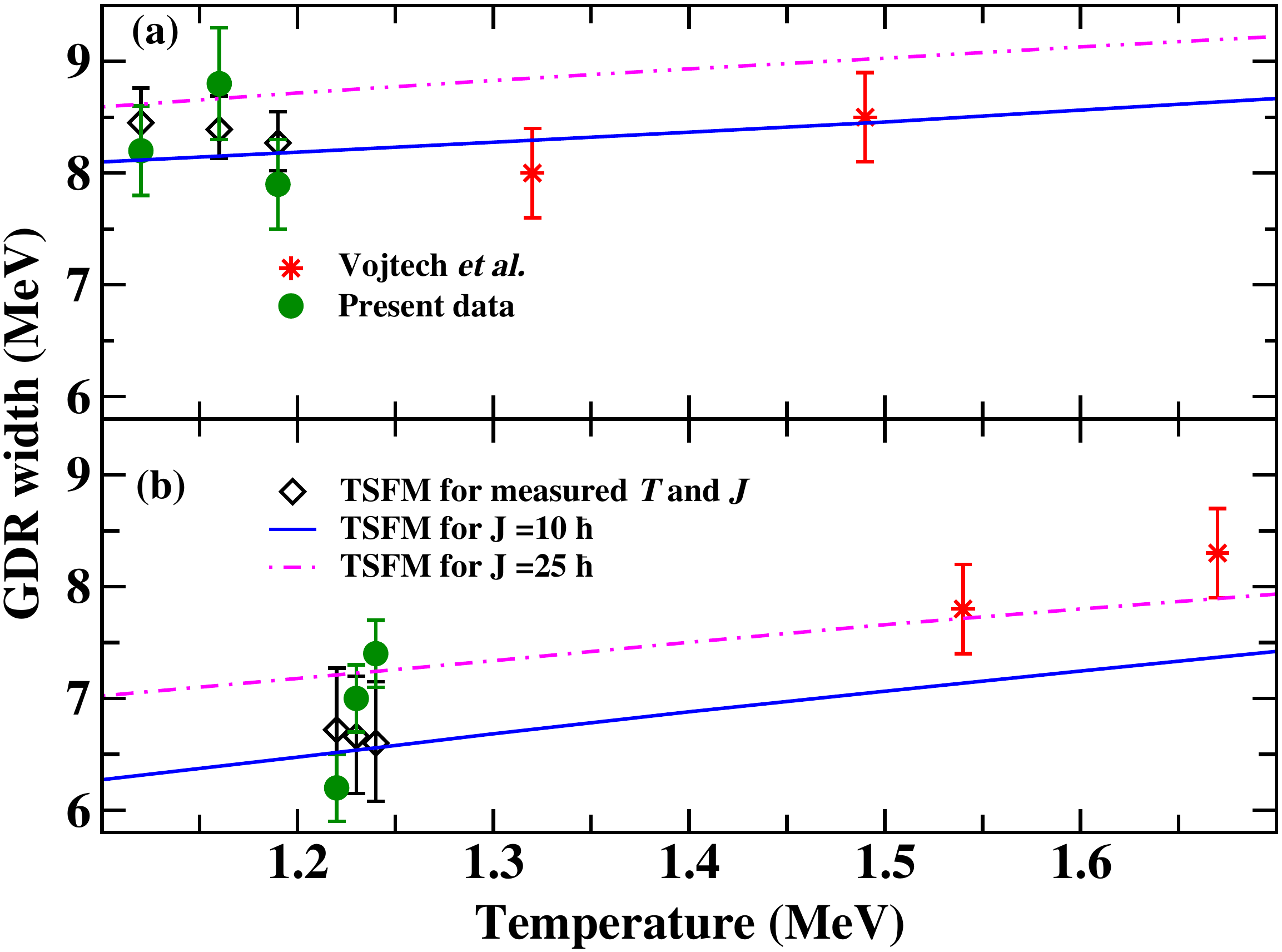}
\caption{\label{widthplot}(Color online) The variation of $\it \Gamma_{\rm GDR}$ with temperature for (a)$\rm ^{124}Ba$ and (b)$\rm ^{136}Ba$ together with $\it \Gamma_{\rm TSFM}$ (open symbols).  The error bars in the $\it \Gamma_{\rm TSFM}$ correspond to the variation resulting from the experimental spread in temperature. The `*' symbols represent the data taken from Vojtech $et$ $al.$~\cite{vojtech}. The prediction of $\it \Gamma_{\rm TSFM}$ for $J$~=~10$\hbar$ (continuous line) and for $J$~=~25$\hbar$ (dashed-dot line), corresponding to angular momentum range spanned by data, are also shown for comparison.}
\end{figure}

The $\it \Gamma_{\rm D}$ for $\rm ^{124}Ba$ ($\rm ^{136}Ba$) from present (filled symbols) and earlier measurements (`*' symbols) are shown as a function of $T$ in the top (bottom) panels of Fig.~\ref{widthplot} together with $\it \Gamma_{\rm TSFM}$ (open symbols). The error bars in the $\it \Gamma_{\rm TSFM}$ correspond to  the variation resulting from the experimental spread in $T$. Within the experimental errors, the observed  $\it \Gamma_{\rm D}$ is in good agreement with the TSFM calculations. This figure also shows $\it \Gamma_{\rm TSFM}$  as a function of $T$ for $J$~=~10$\hbar$ and $J$~=~25$\hbar$, corresponding angular momentum range spanned by the data.
The calculated widths are in reasonable agreement with the data for both the nuclei over the temperature range of 1.1--1.7~MeV. It should be noted that at given $T$-$J$, the $\it \Gamma_D (\rm ^{136}Ba)$ is smaller  than the $\it \Gamma_{\rm D} (\rm ^{124}Ba)$, but the increase in $\it \Gamma_{\rm D}$ with $T$ appears to be more rapid in $\rm ^{136}Ba$ than in the case of $\rm ^{124}Ba$. This is perhaps indicative that the near shell closure effect plays a dominant role at low temperature in $\rm ^{136}Ba$. 
 The advent of radioactive ion beam facilities and modern detection systems will facilitate GDR  studies nuclei with wider $N/Z$ ratio in near future.

\section{Summary and conclusion}
Exclusive measurements of high energy $\gamma$-rays are performed in $\rm ^{124}Ba$ and $\rm ^{136}Ba$ at the same excitation energy ($\sim$~49~MeV), to study properties of the giant dipole resonance (GDR) over a wider $N/Z$ range. The multiplicity of low energy $\gamma$-rays are measured in coincidence with high energy $\gamma$-rays to disentangle the effect of $T$ and $J$. The GDR parameters are extracted employing a simulated Monte Carlo statistical model analysis.
The observed $\gamma$-ray spectra of  $\rm ^{124}Ba$ can be explained with prolate deformation with $<\beta>$~=~0.29, which is very similar to ground state deformation. The observed GDR centroid ($\sim$~16~MeV) and width ($\sim$~8~MeV) remain constant for the studied $T$ (1.12--1.19~MeV) and $J$ (14--22$\hbar$) range. In the case of $\rm ^{136}Ba$, a single component Lorentzian function which indicates spherical shape could explain the $\gamma$-ray spectra very well. In this case also,  no significant variations (beyond experimental errors) are observed in the centroid energy $\sim$~14.7~MeV and the width $\sim$~7.0~MeV. The observed variation in $E_{\rm GDR}$ for $\rm ^{124}Ba$ and $\rm ^{136}Ba$ is consistent with systematics, but the width in the latter is considerably narrower.
For  both $\rm ^{124}Ba$ and $\rm ^{136}Ba$, the width of the GDR is nearly constant in the $T$-$J$ range studied and is in reasonable agreement with the TSFM calculations. Further, it is shown that the variation of $\it \Gamma_{\rm D}$ with $T$ is well reproduced  by the TSFM calculations over the temperature range of 1.1--1.7~MeV, predicting a faster increase for $\rm ^{136}Ba$.
   Further studies for very neutron-rich barium isotopes with upcoming radioactive ion beam facilities would be interesting to probe the variation of GDR properties over a wider range of $N/Z$ asymmetry.

\begin{acknowledgments}
We thank K.V. Divekar, M.E. Sawant, S.C. Nadkar and R.~Kujur for their assistance during setup, D. Pujara, R.D. Turbhekar for target preparation, and the Pelletron Linac Facility staff for smooth operation of the accelerator. The authors are grateful to Dr.~D.R.~Chakrabarty for help with the analysis programs and valuable discussions. A.K.R.K acknowledges RIKEN Supercomputer HOKUSAI GreatWave System, where the numerical calculations were carried out. P.A. acknowledges financial support from the Science and Engineering Research Board (India), SR/FTP/PS-086/2011 and DST/INT/POL/P-09/2014.
\end{acknowledgments}

\end{document}